\documentclass[useAMS,usenatbib]{mn2e}
\usepackage{times, graphicx, setspace, subfigure, latexsym, amssymb, amsmath, fancyhdr}

\title[Multifrequency radio observations of the magnetar PSR\,J1622--4950]{Radio emission evolution, polarimetry and multifrequency single pulse analysis of the radio magnetar PSR\,J1622--4950}
\author[L. Levin et al.]{L. Levin$^{1, 2}$\thanks{E-mail:llevin@astro.swin.edu.au}, 
M. Bailes$^{1,3}$, S. D. Bates$^{4,5}$, N. D. R. Bhat$^{1,3}$, M. Burgay$^6$, S. Burke-Spolaor$^{2,7}$,
\newauthor
N. D'Amico$^6$, S. Johnston$^2$, M. J. Keith$^2$, M. Kramer$^8$, S. Milia$^{6,9}$, A. Possenti$^6$, 
\newauthor
B. Stappers$^5$ and W. van Straten$^{1,3}$\\
$^{1}$Swinburne University of Technology, Centre for Astrophysics and Supercomputing Mail H30, PO Box 218, VIC 3122, Australia\\
$^{2}$Australia Telescope National Facility, CSIRO Astronomy \& Space Science, P.O. Box 76, Epping, NSW 1710, Australia\\
$^3$ARC Centre for All-Sky Astronomy (CAASTRO)\\
$^4$Department of Physics, West Virginia University, 210E Hodges Hall, Morgantown, WV 26506, USA\\
$^5$University of Manchester, Jodrell Bank Centre for Astrophysics, Alan Turing Building, Manchester M13 9PL, UK\\
$^6$INAF-Osservatorio Astronomico di Cagliari, localit\`{a} Poggio dei Pini, strada 54, I-09012 Capoterra, Italy\\
$^7$NASA Jet Propulsion Laboratory, M/S 138-307, Pasadena, CA 91106, USA\\
$^8$Max Planck Institut f\"{u}r Radioastronomie, Auf dem H\"{u}gel 69, 53121 Bonn, Germany\\
$^9$Dipartimento di Fisica, Universit\`{a} degli Studi di Cagliari, Cittadella Universitaria, 09042 Monserrato (CA), Italy}

\begin{document}

\date{Accepted ... Received ...; in original form ...}

\pagerange{\pageref{firstpage}--\pageref{lastpage}} \pubyear{2012}

\maketitle

\begin{abstract}
Here we report on observations of the radio magnetar PSR\,J1622--4950 at frequencies from 1.4 to 17~GHz. We show that although its flux density is varying up to a factor of $\sim$10 within a few days, it has on average decreased by a factor of 2 over the last 700 days. At the same time, timing analysis indicates a trend of decreasing spin-down rate over our entire data set, again of about a factor of 2 over 700 days, but also an erratic variability in the spin-down rate within this time span. 
Integrated pulse profiles are often close to 100 per cent linearly polarized, but large variations in both the profile shape and fractional polarization are regularly observed. Furthermore, the behaviour of the position angle of the linear polarization is very complex - offsets in both the absolute position angle and the phase of the position angle sweep are often seen and the occasional presence of orthogonal mode jumps further complicates the picture. However, model fitting indicates that the magnetic and rotation axes are close to aligned. Finally, a single pulse analysis has been carried out at four observing frequencies, demonstrating that the wide pulse profile is built up of narrow spikes of emission, with widths that scale inversely with observing frequency.
All three of the known radio magnetars seem to have similar characteristics, with highly polarized emission, time-variable flux density and pulse profiles, and with spectral indices close to zero. 
\end{abstract}


\section{Introduction}
PSR\,J1622--4950 was discovered in the High Time Resolution Universe survey for pulsars and fast transients \citep{kei10} currently underway at the Parkes  and Effelsberg radio telescopes. The pulsar's many similarities with the two previously known magnetars that emit radio pulsations have placed this source in the fast growing group of magnetars. Magnetars are commonly thought to be rotating neutron stars that in addition to their emission of pulsating radiation also undergo large bursts and outbursts of radiation in the X-ray and $\gamma$-ray bands (for more detailed reviews on magnetars see e.g. \cite{mer08,rea11}).
The magnetar group is built up of two subgroups: anomalous X-ray pulsars (AXPs) and soft gamma-ray repeaters (SGRs). There is however no longer a strict division between the two classes, as new observations have shown that some of the sources simultaneously exhibit properties originally thought to belong exclusively to only one of the two classes \citep{gav02, mer09, rea09}. 
The term magnetar originates in the sources' extremely high inferred surface magnetic fields (typically $\geq 10^{14}$G) and it is believed that their radiation is powered by the energy stored in the magnetic fields \citep{dun92} instead of by the spin-down as is the case for ordinary pulsars. Recently, a new magnetar (SGR\,0418+5729) with a considerably lower surface magnetic field ($B < 7.5\times 10^{12}$G) was discovered \citep{rea10}, casting doubts on the assumption that a high surface dipolar magnetic field strength is a requirement for magnetar-like activity. 

Radio pulsations from a magnetar were first detected in 2006 from the source XTE\,J1810--197 \citep{CamNature06}, and since then only two other sources have been found to belong to the group of radio-emitting magnetars: 1E\,1547.0--5408 \citep{CamApJ666} and PSR\,J1622--4950 \citep{lev10}. 
XTE\,J1810--197 and 1E\,1547.0--5408 are both so-called transient magnetars, that occasionally undergo large outbursts of X-ray emission. The radio properties of these two pulsars have been described in detail in a series of papers \citep[e.g.][]{CamApJ659, CamApJ679, kra07, laz08}, which reported on features that make the radio magnetars stand out from the ordinary pulsar population. 
In addition to long pulse periods and high surface magnetic field strengths, these features include highly variable radio flux densities, changing pulse profiles on short time scales, large amounts of timing noise and a flat radio spectrum. 

Both sources emit nearly 100\% linearly polarized radiation at a large range of observing frequencies \citep{kra07,CamApJ659,CamApJ679}. Analyses of the linear polarization position angle (PA) show a preferred neutron star geometry for 1E\,1547.0--5408 where the rotation and magnetic axes are close to aligned \citep{CamApJ679}, which at the time was supported by a low pulsed fraction in the X-ray \citep{hal08}. Since then, X-ray monitoring has shown much higher pulse fraction values \citep{isr10} and it has been shown that the low pulsed fraction in X-ray observed during high flux states may be due to a dust scattering halo \citep{isr10,ber11,sch11}. This is in conflict with the aligned geometry derived from the radio emission of 1E\,1547.0--5408.
The radio emission geometry analysis for XTE\,J1810--197 has proven difficult and different groups have reported on different results. \cite{CamApJ659} report on two possible solutions for the geometry: Either the magnetic and rotation axes are nearly aligned, or the emission originates high above the surface of the star. \cite{kra07} on the other hand derive a geometry where two emission cones must be present in the neutron star magnetic field. This is interpreted as either an offset dipole or a non-dipolar field configuration. Analyses of the X-ray data from XTE\,J1810--197 seem to favor a non-aligned geometry for this source \citep{per08}.

Single pulse studies of XTE\,J1810--197 are covered in great detail in \cite{ser09}. They show that the integrated pulse profile consists of strong spiky sub-pulses, with an overall high modulation index that varies between components of the pulse. Their analysis concludes that the radio emission from XTE\,J1810--197 is clearly different to that from ordinary pulsars.

The radio magnetar PSR\,J1622--4950 was discovered by \cite{lev10}. 
That paper reports on the high variability of this pulsar in the shape of the integrated pulse profile and in radio flux density on time scales of hours. The X-ray counterpart of the pulsar is identified, with an observed X-ray luminosity $L_{\rm X}$(0.3--10 keV) $ \approx 2.5 \times 10^{33}$\,ergs\,s$^{-1}$, which is in the lower end of the range of X-ray luminosities observed for quiescent magnetars ($1.8 \times 10^{33}<L_{\rm X}$(1--10keV) $<1.2 \times 10^{36}$; \cite{rea11}; or see the McGill SGR/AXP Online Catalogue\footnote{http://www.physics.mcgill.ca/$\sim$pulsar/magnetar/main.html}).
The same paper also mentions that PSR\,J1622--4950 has highly linearly polarized radio emission and an inverted radio spectrum. Further studies of the radio spectrum of this magnetar have been carried out by \cite{kei11}, who observed it at bands centered at 17\,GHz and 24\,GHz, and concluded that the spectral index is close to zero when the flux densities from these observations are added in the radio spectrum calculation. 

In this paper, we will present the continued observations and analysis carried out for PSR\,J1622--4950 with the Parkes radio telescope. We will focus on four different aspects of the emission:
first we will have a look at the flux density evolution over $\sim$2 years of observations to see if the source is still as variable in flux density as has been measured previously and if there are any trends in how it is varying overall.
The second part will treat the timing analysis of the source, with the complications that a highly varying pulse profile introduces to this process. We will investigate if the frequency ($\nu$) and frequency derivative ($\dot{\nu}$) that we observe demonstrate the true spin-down of the pulsar or if their changes are artifacts of the pulse profile variations. 
Thereafter we will report on the polarimetry of the integrated pulse profiles and compare the results to the two previously known radio magnetars. We will also attempt to derive the geometry of the radio emitting regions of the neutron star.
Finally we will report on an analysis of single pulses from the magnetar at several observing frequencies. Throughout the paper we will review the similarities and differences between PSR\,J1622--4950 and 1E\,1547.0--5408, XTE\,J1810--197 and ordinary pulsars. 
This discussion will be incorporated in each pulsar property section in the paper, and we will finish by summarizing our findings in the last section.

\begin{figure*}
	\begin{center}
		\includegraphics[width=15cm]{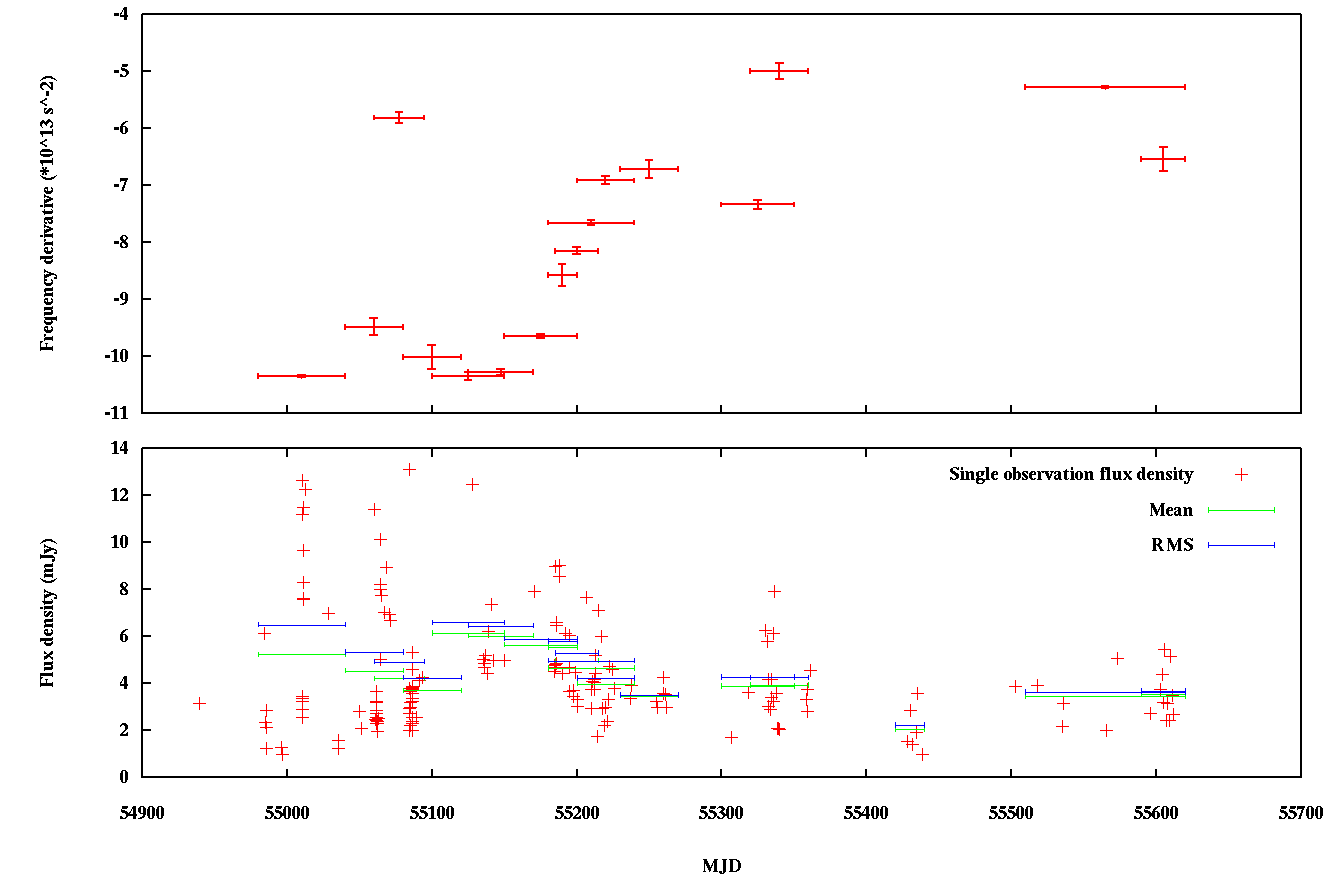}
	\end{center}
	\caption{{\it Top plot:} Variation of frequency derivative with time. The error bars in the x-range indicate the time span of the included TOAs. The large variations in the value of the frequency derivative indicate changes in absolute value and sometimes also in sign of the second derivative. In some cases we have included a fit to a shorter data set, even if all the TOAs in that set are already included in a fit of a longer data set, to show the direction of the frequency derivative change. In total, the values of the frequency derivative varies up to a factor of $\sim$2. 
	{\it Bottom plot:} Variation of flux density at 1.4 GHz with time. The mean and RMS values were calculated using the same sets of observations as were used in the frequency derivative analysis.} 
	\label{Fig:fdot-mjd}
\end{figure*}


\section{Observations and Analysis}
All data used for this analysis were collected with the 64-m dish at the Parkes Radio Telescope using different receivers and backends. Observations were made at frequency bands centered at: 1.4\,GHz using the center beam of the Multibeam Receiver \citep{sta96}, 3.1\,GHz using the `10-/50-cm' receiver and 17 and 24\,GHz using the `13-mm' receiver.

The Parkes Digital Filterbank System (PDFB3) used to create the folded profiles first converts the analogue voltages from each polarization channel of the linear feeds into digital signals. It then produces 1024 polyphase filterbank frequency channels that are folded at the apparent topocentric period of the pulsar into 1024 pulsar phase bins, and written to disk every 20\,s. Four Stokes parameters are recorded. To determine the relative gain of the two polarization channels and the phase between them, a calibration signal is injected at an angle of 45$^\circ$ to the feed probes. 
The data are analysed off-line using the {\sc psrchive} package\footnote{See http://psrchive.sourceforge.net} \citep{hot04} and corrected for parallactic angle and the orientation of the feed. The position angles are also corrected for Faraday rotation through the interstellar medium using the nominal rotation measure.

For the single pulse analysis the baseband data recording and 
processing system known as the ATNF Parkes Swinburne Recorder \citep[APSR;][]{van11} was used.


\section{Radio Light Curve}

As reported in \cite{lev10} the 1.4\,GHz flux density of the integrated pulse profile varies greatly between observations. Since then we have collected about one more year of data on this source. Plotted in the bottom panel of Fig. \ref{Fig:fdot-mjd} is the integrated flux density of each observation made at 1.4\,GHz at Parkes since the discovery in April 2009. The new data points begin around MJD=55240. It is clear from this plot that the peak observed flux density during the last $\sim$200 days is only about half the value observed during the first $\sim$200 days after the discovery, suggesting an intrinsic long-term decay of the flux density. However, it is possible that the magnetar is still just as variable and with as high peak flux density as before but that the higher flux density points are missed during the later time span due to the smaller sample. Simply fitting a line to the data points results in a slight slope, giving a decline of the average flux density of $\sim$2\,mJy for the 700 days of observing. 
To more easily visualise the flux density decline we have divided the data points into smaller sets of about 30 days each and calculated the mean flux density and the root mean square (RMS) for each of these data sets. 
The time span for each data set is the same as we used for calculating different rotation frequency derivatives as described in Sec. \ref{Sec:timing} below. 
The mean and RMS of each data set is plotted on top of the data points in the bottom panel of Fig. \ref{Fig:fdot-mjd}.

\cite{and11b} report on recent observations of PSR\,J1622--4950 with the Chandra X-ray Observatory and the Australia Telescope Compact Array (ATCA) that were collected within the framework for the "ChIcAGO" project \citep{and11a}. They observed the magnetar with the ATCA simultaneously at frequency bands centered at 5.0 GHz and 9.0 GHz on November 22, 2008 and December 5, 2008. The flux densities are 33.0$\pm$0.3\,mJy  and 40.4$\pm$0.3\,mJy at 5.0\,GHz and 30.9$\pm$0.6\,mJy and 31.9$\pm$0.6\,mJy at 9.0\,GHz for the two observation sessions respectively. This is significantly higher flux densities than the values measured on December 8, 2009 and February 27, 2010 with the ATCA published in \cite{lev10}: 13$\pm1$\,mJy at 5.0\,GHz and 14.3$\pm$0.8\,mJy at 9.0\,GHz. These values indicate a decrease in flux density of about 68\% at 5.0\,GHz and 55\% at 9.0\,GHz over one year \citep{and11b}. With this in mind it seems likely that the flux density decline observed recently at 1.4\,GHz with Parkes is a real intrinsic decay of flux density in the magnetar. 


\section{Timing}
\label{Sec:timing}

\begin{figure}
	\begin{center}
		\includegraphics*[width=8cm]{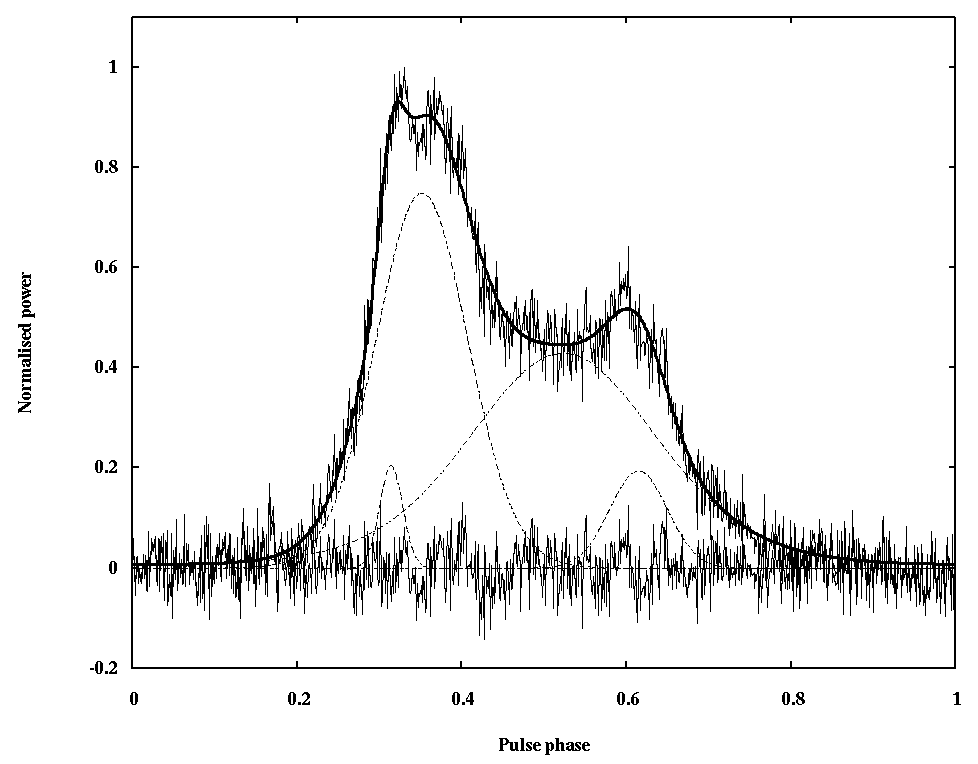}
	\end{center}
	\caption{Timing model. The dashed lines show the von Mises functions used to build up the profile, which added together give the solid line that follows the pulse shape. The bottom noise level shows the emission that is left after subtracting the model. Profile from 29 June 2009.}
	\label{Fig:model}
\end{figure}

To analyse the rotational history of a pulsar, it is conventional to integrate over a set number of rotations, enough to create a stable pulse profile for each observation. These profiles are then aligned with the help of a standard profile, unique to that particular pulsar, to create a list of the pulse times of arrival (TOAs) at the telescope. The list of TOAs is used to determine a more accurate period and spin-down rate of the pulsar as well as its precise position in the sky, a more precise value of its dispersion measure (DM), etc. When performed on data from a large enough time span, using this method generally results in values for the pulsar parameters with very high accuracy.

However, due to the variability of pulse profile at 1.4 GHz, the timing analysis of PSR\,J1622--4950 is more complex than for an ordinary pulsar. Here we have employed the same tools in the timing procedure as were used in \cite{lev10}, i.e. TOAs for the pulses were calculated by using a model that describes the different components of the profile, rather than using a standard profile. 
The model was created using software from the {\sc psrchive} package \citep{hot04} by fitting scaled von Mises functions \citep{mis18} to the pulse profile of one of the observations where all the components were present. This profile and the corresponding model is shown in Fig. \ref{Fig:model}. The TOAs are then created by letting the amplitudes of the components vary but keeping the separations fixed, while fitting the model to each observation. The timing analysis was made using the {\sc Tempo}
software\footnote{See http://www.atnf.csiro.au/research/pulsar/tempo/}. 

However, we cannot obtain a coherent timing solution for our full data span even using this method.
Instead we have looked at data from shorter time spans and fitted the observed frequency ($\nu$) and frequency derivative ($\dot{\nu}$) for each set of timing points separately, in an attempt to quantify how much and in which direction the true spin-down is changing with time. The number of points included in each set is dependent on how long we could get a stable $\dot{\nu}$ with a reasonable error 
(largest error in Fig. \ref{Fig:fdot-mjd} is $\dot{\nu}_{err}$ = 0.21 $\times$ 10$^{-13}$\,s$^{-2}$). 
For this timing analysis the position is held fixed at that constrained by the X-ray counterpart: R.A. = 16:22:44.80, Dec. = -49:50:54.4 \citep{lev10} and the DM is set to 820 cm$^{-3}$pc.
The result of this analysis is shown in the upper panel of Fig \ref{Fig:fdot-mjd}. It is evident from this plot that $\dot{\nu}$ has been changing with a factor of $\sim$2 since the discovery, as was also stated in \cite{lev10}. 
However, the second derivative of the frequency ($\ddot{\nu}$) is also changing rapidly in magnitude and direction with time.

A comparison of these results to the timing analysis carried out previously for the other two radio magnetars, 1E\,1547.0--5408 \citep{CamApJ663} and XTE\,J1810--197 \citep{CamApJ679} shows clear similarities, but also differences. 
In all three cases the $\dot{\nu}$ is changing greatly as a function of time. In both of the other sources the $\ddot{\nu}$ seems to vary more smoothly than it does for PSR\,J1622--4950 and they both have steady trends along a fairly straight line (allowing for some "wobbling" on the way). However, the sign of the estimated $\ddot{\nu}$ is different for the two sources: XTE\,J1810-197 has a positive $\ddot{\nu}$ and 1E\,1547.0--5408 has a negative one. 
The timing analyses for both the other two magnetars were made using data that were collected more regularly and frequently than our observations, which may contribute to the smoother looking frequency derivative evolution. 

\cite{CamApJ663} quantify the timing noise by looking at the magnitude of the cubic term of a Taylor series expansion of rotational phase over a time interval $t$, i.e. $\ddot{\nu}$\,$t^{3}/6$ \citep{arz94}. By using this expression they get about 60 cycles over 6 months for 1E\,1547.0--5408 and 20 cycles over 9 months for XTE\,J1810--197. The same calculation for PSR\,J1622--4950 yields about 250 cycles over the entire 20 months that the pulsar has been observed since the discovery. If we instead look at only the 100 days when the $\dot{\nu}$ is steadily increasing (from MJD$\sim$55150 to MJD$\sim$55250) the value is about 5 cycles over 3.3 months. 

One option that could explain the unusual behavior of $\dot{\nu}$ in PSR\,J1622--4950, is if the source went through a glitch shortly before the discovery observation at MJD=54939. Glitches are more frequent for younger pulsars than for older sources, but in general unusual in ordinary pulsars \citep{esp11}. For AXPs however, glitches have been observed in nearly all known sources \citep{dib08} and it would therefore be feasible to observe a glitch also for PSR\,J1622--4950.
Unfortunately it is very hard to constrain if a glitch has occurred without any data from around or before the time of the possible glitch, but the possibility is worth noting for future timing efforts of this source. 

By comparing the two panels in Fig. \ref{Fig:fdot-mjd}, there is an apparent correlation between reduced flux density and increased $\dot{\nu}$, again especially between MJD $\approx$ 55100 and MJD $\approx$ 55300. Calculating Pearson's correlation coefficient, $r_{(x,y)}$\,=\,cov(x,y)/$\sigma_x\sigma_y$ (where cov(x,y) is the covariance of x and y, and $\sigma$ is the standard deviation), for correlation between $\dot{\nu}$ and the mean values for each time span in Fig.  \ref{Fig:fdot-mjd} gives $r$\,=\,--0.73, which corresponds to a 3 sigma correlation. It is however hard to constrain that these two features are direct consequences of each other, and we stress that care should be taken not to infer too much from these correlation results. 

\cite{bel09} describes a model in which starquakes or glitches in the neutron
star cause the magnetic field lines to twist. The author describes the
electrodynamics of the untwisting of the field lines as they relax back
towards their initial state. He considers this model in the context
of the magnetars, in particular XTE\,J1810--197, where the reduction in the radio
flux density over time, the changes to the torque and the X-ray observations
all conform roughly with his calculations. This model may also be
applicable here as we see a similar decrease in flux density over time and
large variations to the torque although we are hampered by a lack of
information in the X-ray band. It is also hard to see how the gradual
untwisting of magnetic field lines can cause the rapid profile and
polarization variations that we see as these appear to oscillate back
and forwards between states on a time-scale much shorter than the decay of the radio flux density.


\section{Polarimetry}

\begin{figure*}
	\begin{center}
		\includegraphics[width=7cm]{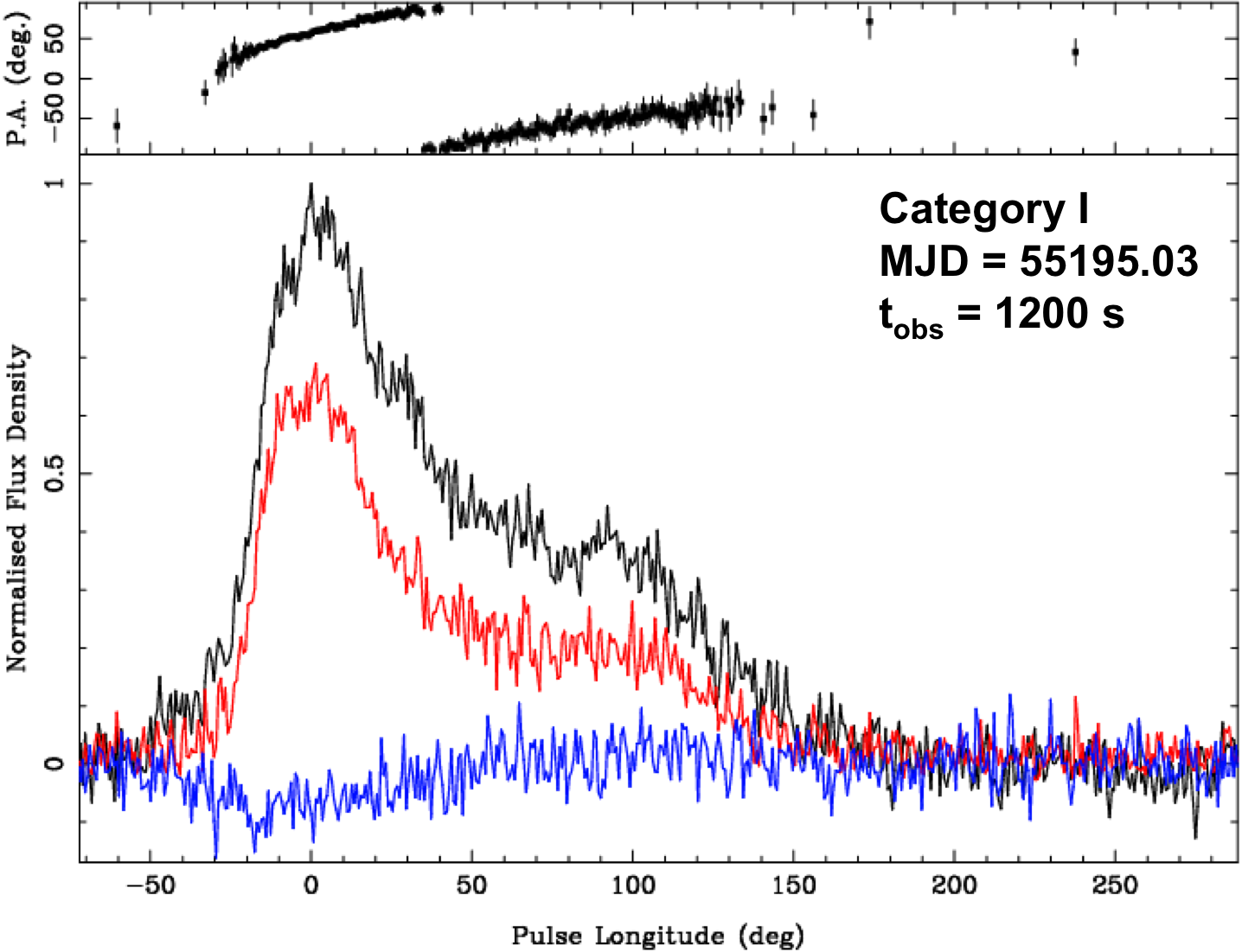}
		\includegraphics[width=7cm]{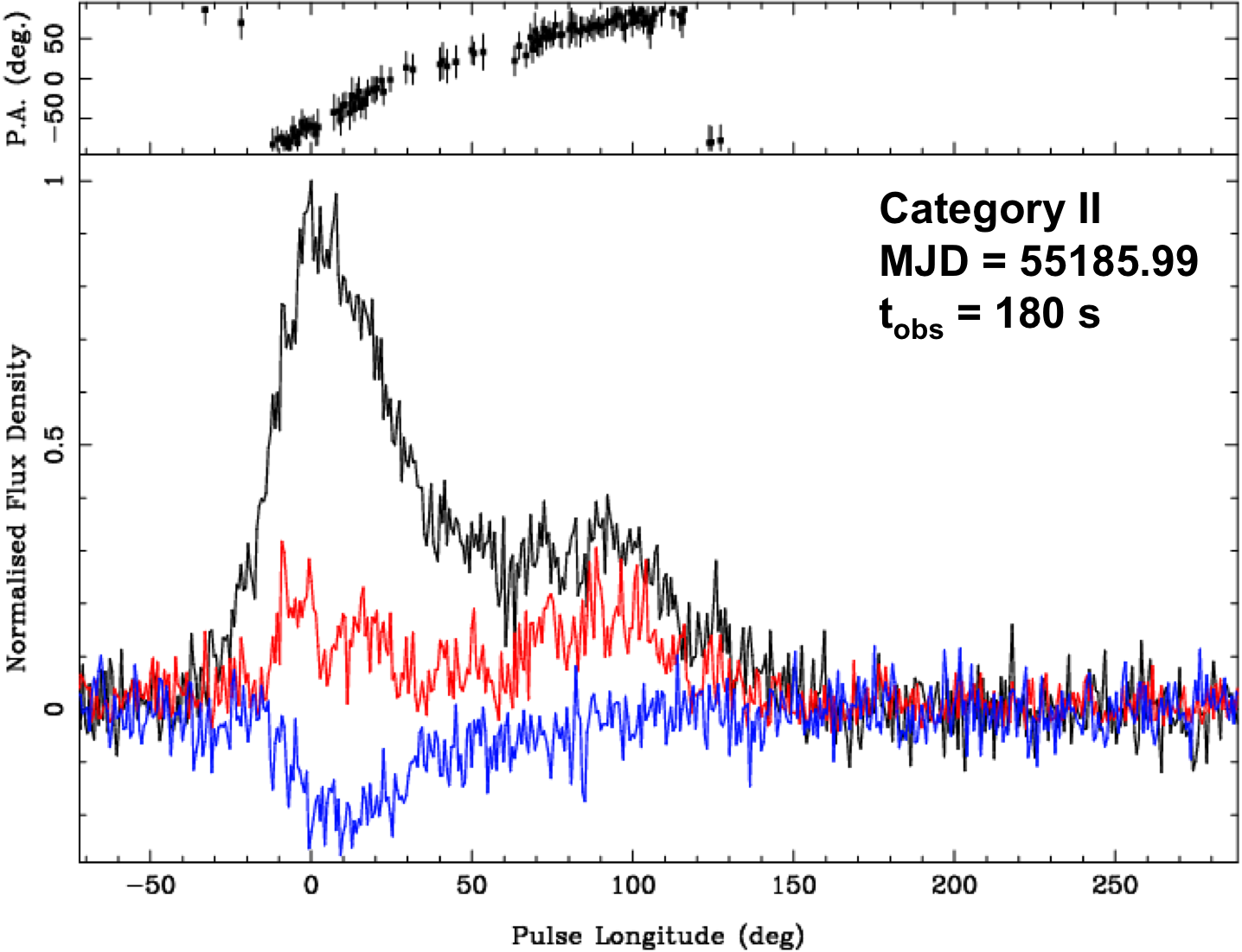}
		\includegraphics[width=7cm]{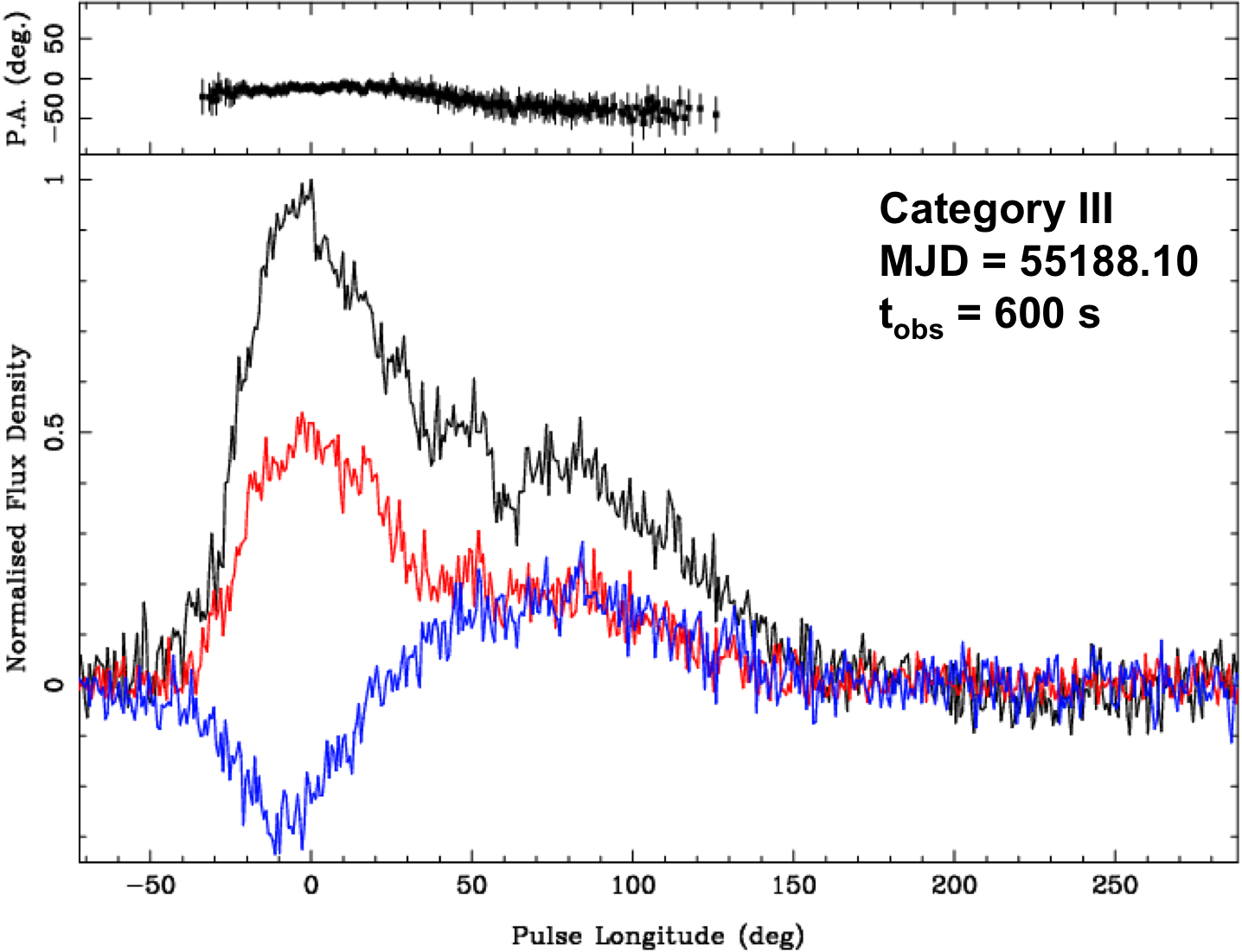}
		\includegraphics[width=7cm]{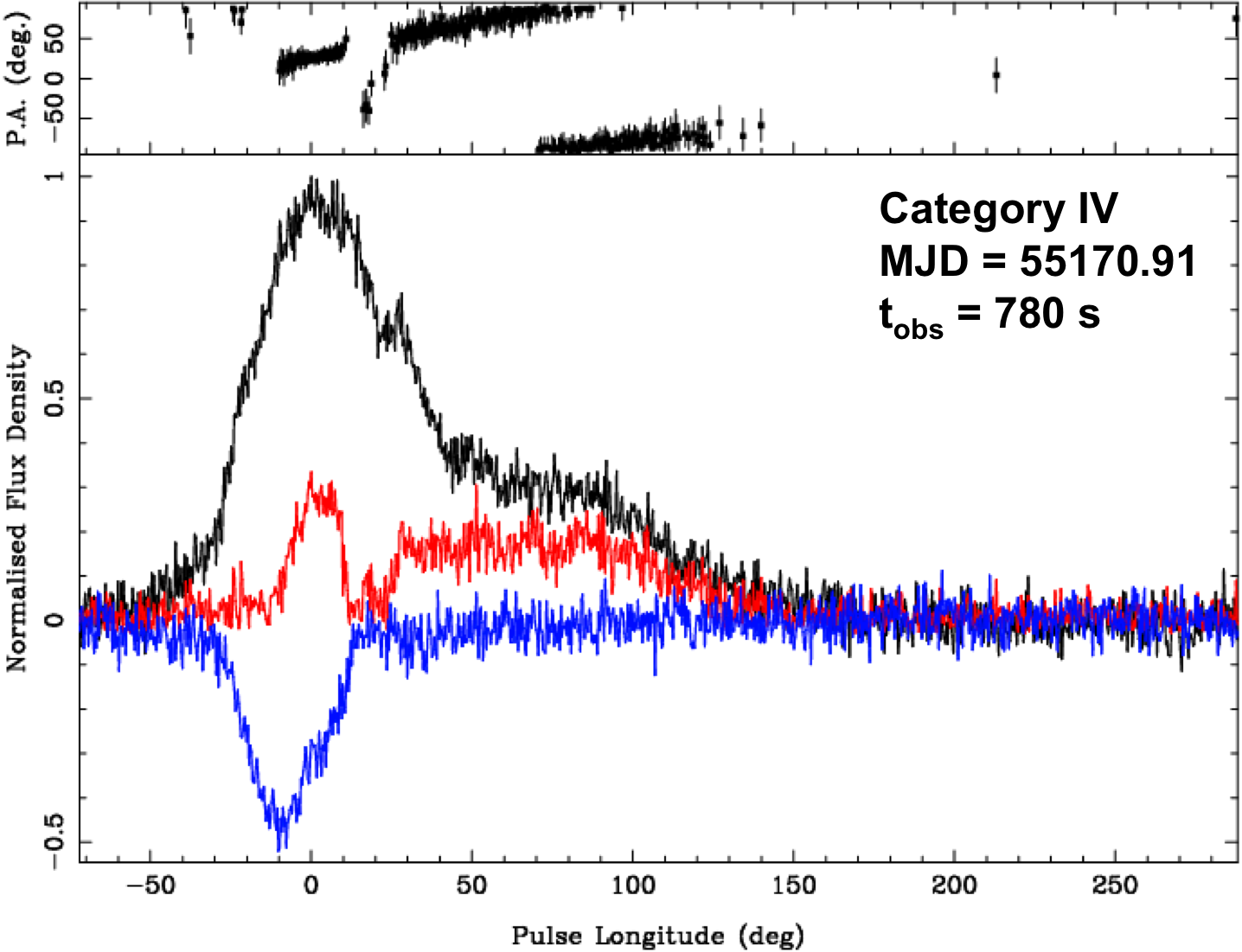}
	\end{center}
	\caption{Examples of observations at 1.4\,GHz from the four different polarization groups described in Sec \ref{sec:polcat}. Category I - IV are shown from top left to bottom right. Total intensity is shown in black, linear polarization in red and circular polarization in blue. The integration time for each observation is noted in the panels, however there are both shorter ($\sim$180\,s) and longer ($\sim$600\,s) observations present in all categories.
	Note the short time scale of the variations: all four observations were collected within the same calender month (December 2009).}
	\label{Fig:polgroups}
\end{figure*}

We have collected polarimetric data for PSR\,J1622--4950 with the Parkes telescope at four different observing frequencies, with bands centered at: 1.4\,GHz (the centre beam of the Multibeam receiver), 3.1\,GHz (10-/50-cm receiver) and 17 and 24\,GHz (13\,mm receiver). 

The high frequency observations (at 17 and 24\,GHz) were reported on in \cite{kei11}, and we note some striking similarities in the polarized emission over all observed frequencies. As well as having a very wide ($\sim$50\%) duty cycle for the integrated profiles, the linear polarization is often close to 100\% of the total intensity. However, similar to the flux density and pulse profile shape, also the polarization of the integrated profiles is changing between observations.

\subsection{Polarisation categories}
\label{sec:polcat}
In an attempt to get an overview of the way the polarization is varying we have collected observations with similar characteristics in groups, giving us four separate categories with different properties.
Category I is the most frequent mode, with almost half (48\%) of all categorised observations. It collects the observations where the linear polarization is $>50\%$ of the total intensity, the circular polarization is low and the position angle (PA) has a steep and consistent swing. 
In category II the linear polarization is much lower throughout the profile. In the cases where a second component is present, the fraction of linear polarization is higher in the trailing edge than in the leading edge of the profile. Similar to category I, the circular polarization is low and the PA swing is steep and consistent. 24\% of our observations belong to this group.
For category III (11\% of the observations) the most prominent feature is the shallow PA curve. In addition the profiles also tend to have low linear polarization in the leading edge of the profile and a higher value of the circular polarization than for categories I and II. 
The last 17\% of the observations do not fit into any of the first three groups, and hence will end up in category IV. Here we have collected the observations with jumps and other irregularities in the PA swing, change of handedness in the circular polarization and large changes in linear polarization within the pulse profile. 
Examples from all four groups are shown in Fig \ref{Fig:polgroups}.

When looking at the time evolution of the polarized emission, by analysing observations from the different groups in time order, it does not seem like the variations are following any preferred order, but are fairly random in time. 
The large variations in linear polarization and PA for PSR\,J1622--4950 are at odds with what is seen for XTE\,J1810--197 by \cite{kra07}. They observe an evolution in PA swing over a time-scale of weeks, but very few changes on shorter time-scales. Also \cite{CamApJ659} observed that the general polarization properties of XTE\,J1810--197 do not seem to vary with time as the total intensity changes. That is, the linear polarization is always close to 100\% of the total intensity, and the circular polarization component is low. 
This suggests that the observed profile variations are not caused by changes in the magnetic field geometry of the emission regions for this source. 
Also in the case of 1E\,1547.0--5408 there is little variation in linear polarization and in measured PA swing with observing frequency and time \citep{CamApJ679}. The circular polarization however increases with decreasing frequency, and is overall higher than for XTE\,J1810--197.

\subsection{Rotating Vector Model predictions on the neutron star geometry}

\begin{table*}
\begin{center}
\caption{RVM fits to 3.1\,GHz data. All angles are given in degrees.}
\begin{tabular}{c c c c c c c}
 Obs ID & MJD & $\chi^2$ & $\psi_0$ & $\zeta$ & $\alpha$ & $\phi_0$\\
 \hline
 \hline
s091230\_210826 & 55195.88 & 1.47 & -5.7$\pm$13.6 & 15.1$\pm$28.8 & 36.0$\pm$60.2 & 141.6$\pm$9.7\\
s091230\_215432 & 55195.91 & 1.28 & -11.6$\pm$3.4 & 15.5$\pm$9.8 & 28.6$\pm$17.2 & 137.5$\pm$2.0\\
s100101\_201716 & 55197.85 & 1.01 & -20.6$\pm$2.8 & 20.7$\pm$4.0 & 46.0$\pm$7.8 & 149.7$\pm$1.6\\
t100615\_074911 & 55362.33 & 1.34 & -20.5$\pm$1.8 & 14.7$\pm$5.8 & 28.3$\pm$10.4 & 137.0$\pm$1.1\\
t100825\_102356 & 55433.43 & 1.62 & -23.7$\pm$6.1 & 16.3$\pm$22.2 & 25.0$\pm$33.5 & 157.4$\pm$2.2\\
t110116\_211759$^{a}$ & 55577.89 & 2.40 & -4.8$\pm$1.5 & 13.2$\pm$7.0 & 22.2$\pm$11.3 & 192.9$\pm$0.9\\
s110410\_210120 & 55661.88 & 1.69 & -29.6$\pm$6.4 & 7.1$\pm$21.9 & 20.7$\pm$58.0 & 122.6$\pm$7.3\\
s110410\_212455 & 55661.89 & 1.10 & -26.5$\pm$3.2 & 9.9$\pm$9.7 & 24.7$\pm$21.8 & 128.9$\pm$3.3\\
\hline
\multicolumn{7}{l}{Notes:}\\
\multicolumn{7}{l}{$^{a}$ Observation with orthogonal PA jump}\\
\end{tabular}
\label{tab:rvm}
\end{center}
\end{table*}

By analysing the linear polarization and how its position angle is varying across the pulse profile, predictions on the angles of the rotation and dipole axes can be made.
The rotating vector model \citep[RVM;][]{rad69} states that the pulsar emission beam has its base close to the dipole axis of the pulsar magnetic field and is observed through rapid swings of the linear polarization position angle over the pulse phase. 
When using the RVM there are a number of effects that are very difficult to take into account and hence are often ignored. These effects include rotational sweepback of the magnetic field lines \citep{dyk04}, propagation effects in the pulsar magnetosphere \citep[e.g.][]{pet06}, emission height differences \citep{dyk08} and multipolar components of the magnetic field. 
There are a number of papers discussing these effects on the magnetic field geometry for magnetars in particular, and the possibility that higher order multipoles are contributing to the magnetic field structure near the magnetar surface \citep[e.g.][]{tho02,bel09,rea10,tur11}. We will disregard these effects in this paper, but it is important to keep in mind that the magnetic field topology may well deviate from the simple dipole model.

In the case of PSR\,J1622--4950, the highly varying values of the PAs in the 1.4\,GHz observations make it difficult to find a consistent solution to the geometry of the neutron star emission. RVM fits to data from different days give different answers depending on the parameters of the PA curve for that particular observation. 
The scatter broadening of the single pulses at 1.4\,GHz (as described in section \ref{Sec:SPchar}) could contribute to some of the large changes in linear polarization and PA swing that we observe at this observing frequency, and hence we have focused the RVM fits to the less scatter broadened and apparently more stable 3.1\,GHz data. 
Tab. \ref{tab:rvm} shows the values from the 3.1\,GHz RVM fits. Even though the best fit angles vary between observations we are able to put some constraints from these fits. The angle between the spin axis and the pulsar-observer line of sight, $\zeta$, is small in all observations with $\zeta \lesssim 20^\circ$ and the angle between the spin axis and the magnetic pole, $\alpha$, is always just a few degrees higher than $\zeta$ with $20^\circ \leq \alpha \leq 46^\circ$. The resulting angle, $\beta$, places the line of sight between the magnetic pole axis and the positive rotation axis with values $-25^\circ \leq \beta \leq -8^\circ$.
The results imply that the pulsar has close to aligned magnetic and rotation axes and that the line of sight remains within the emission beam for large parts of the rotation, which in turn is implied by the wide duty cycle of the integrated profiles. 

RVM predictions for 1E\,1547.0--5408 are described in \cite{CamApJ679}. They carried out polarimetric studies with several different receivers at the Parkes telescope at five different observing frequencies ranging from 1.4 to 8.4\,GHz and at the Australia Telescope Compact Array (ATCA) at frequency bands centered at 18 and 44\,GHz. They report on a slow PA sweep that has an absolute value identical at all observed frequencies, and their RVM fit suggests nearly aligned rotation and magnetic axes. At the time, this result was strengthened by a low pulsed fraction in the X-ray ($\sim$7\%) measured as the source was in quiescence, shortly before the 2007 outburst \citep{hal08}. It has subsequently been suggested that the low pulsed fraction may be due to a dust scattering halo around the magnetar \citep{tie10,ola11}.
Observations of the magnetar at the time of the 2008 October outburst showed much higher pulsed fraction values ($\sim$20\%) which increased to $\sim$50\% during the following few weeks \citep{isr10}.
The anti-correlation between the flux and the pulsed fraction that has been observed for 1E\,1547.0--5408 \citep{isr10,ber11,sch11} together with the highly variable pulse profiles as a function of time, may suggest that the magnetospheric geometry is variable during the initial phases after an outburst and may not be easily related to the geometry in quiescence \citep{isr10}. Hence, even though the higher pulsed fraction measured for 1E\,1547.0--5408 in more recent observations conflict with the aligned geometry derived from the radio data, we do not see these results as strong enough to completely rule out an aligned geometry.

Analyses of XTE\,J1810--197 also result in different geometries for different research groups and wavelengths. 
\cite{CamApJ659} analysed radio polarization observations of XTE\,J1810--197 collected with the Parkes Telescope at three observing frequencies (1.4, 3.2 and 8.4\,GHz). Their analysis shows a shallow swing of the position angle of the linear polarization, with values that yield two possible solutions for the geometry of the magnetar. Either the magnetic and rotation axes are nearly aligned, or the emission originates high above the surface of the star. 
Using X-ray data, \cite{per08} also estimated the viewing geometry of XTE\,J1810--197. 
They determine the allowed minimum and maximum angles between the line of sight and the emission hot spot, and find that the range of the minimum value is compatible with very small angles (including zero) while the maximum is always large ($\gtrsim 60^{\circ}$). 
It has been shown that the peaks of the radio and X-ray pulses from XTE\,J1810--197 are well matched \citep{CamApJ663}, which suggests that the radio emission axis and the hot spot axis are very close to aligned. \cite{per08} found that the high emission height solution from \cite{CamApJ659} was well compatible with their results and that an aligned geometry was unlikely. However, they were not able to make a formal statistical comparison with the results from \cite{CamApJ659}. 
Concurrently, \cite{kra07} also published results from radio polarization observations of the same magnetar. Their simultaneous multifrequency  observations (at 1.4, 4.9 and 8.4\,GHz) were done with three European telescopes: the 76-m Lovell radio telescope in the UK, the 94-m equivalent Westerbork Synthesis Telescope in the Netherlands, and the 100-m radio telescope at Effelsberg in Germany. The main difference in their results compared to \cite{CamApJ659} is that \cite{kra07} include studies of the single pulse emission and identify an inter-pulse in addition to the main pulse during some of their observations and at some frequencies, that was not visible in the Parkes data. The PA values in the inter-pulse are observed to vary with time, which complicates the geometry analysis. Instead of fitting a RVM to the entire pulse profile the authors analysed the main pulse and the inter-pulse separately, with a geometry where two emission cones must be present in the neutron star magnetic field as result. This is interpreted as either an offset dipole or a non-dipolar field configuration. 

The many different geometry analyses that have been carried out for the three radio magnetars make a comparison between them complicated. It is still a bit unclear but we can not completely rule out that the three sources are all aligned, as hinted by the RVM. 
We find this solution very tempting, as it could prove insight in the radio behavior of the magnetars.  
Since the probability of observing a pulsar is lower for a smaller $\alpha$, aligned geometries of the radio magnetars could explain why only three of the 23 currently known magnetars and magnetar candidates have observed radio pulsations. 
So far no X-ray pulsations have been observed from PSR\,J1622--4950. Detection of pulsations in the X-ray might provide additional constraints on the geometry of PSR\,J1622--4950 and would help us to investigate this question further.

\begin{figure*}
	\begin{center}
		\includegraphics[width=12cm]{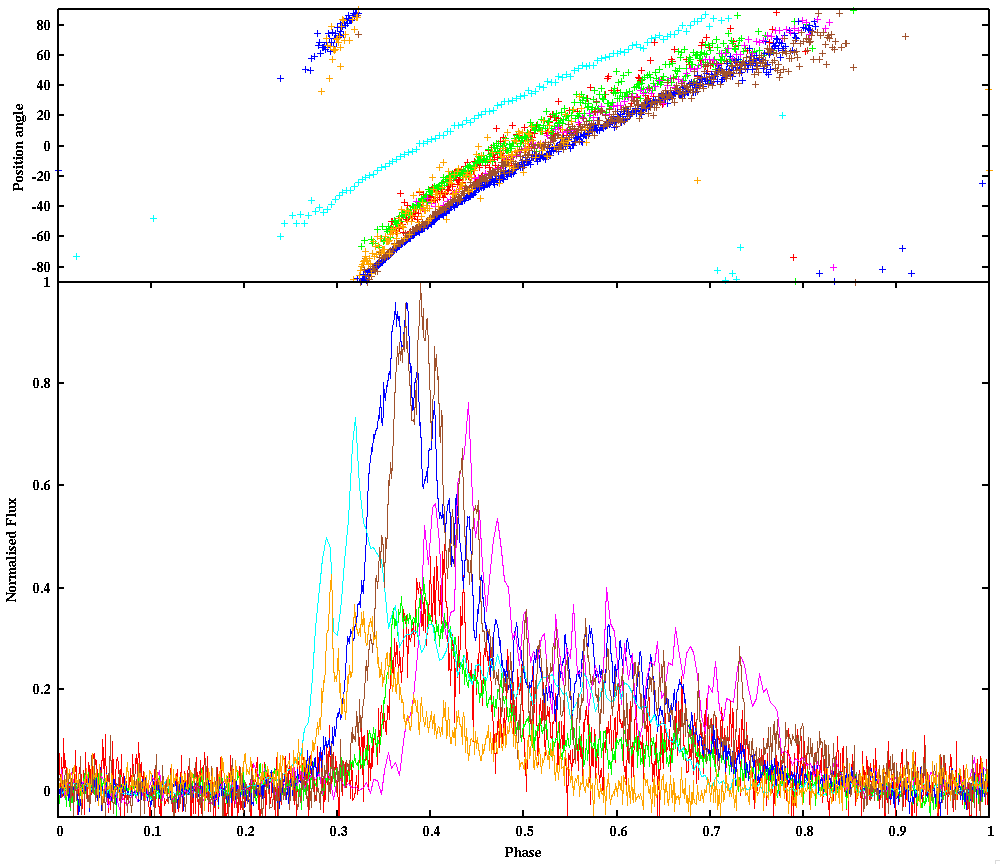}
		\includegraphics[width=12cm]{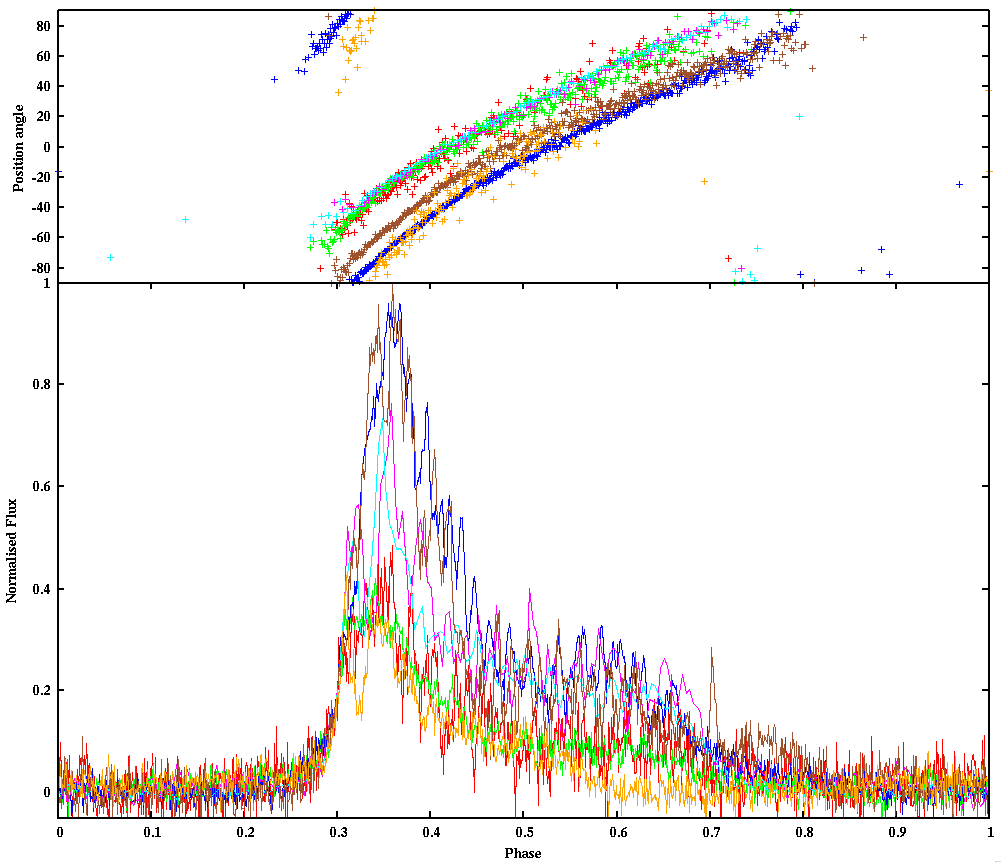}
	\caption{Pulse profiles and PAs for 3.1\,GHz observations without orthogonal PA jumps. The flux density in the lower panels of the two plots are normalised after the maximum flux density of the brightest observation. The profiles of the same color refer to the same observation for both alignments. {\it Top:} The PAs are aligned after the value of $\phi_0$ from the fit to the RVM model. Hence the maximum value of $\delta \psi/\delta \phi$ for each observation is centered at phase 0.3 {\it Bottom:} The profiles are aligned after the total intensity by centering the leading edge of each profile at phase 0.3.}
	\label{Fig:align}
	\end{center}
\end{figure*}

\subsection{Pulse profile alignment}
At 3.1\,GHz the amount of linear polarization is in general very large. All of our observations at this frequency band belong to category I or category IV in the classification above. Even though the amount of linear polarization seems fairly constant, the integrated pulse profile still has a varying shape. The problem with aligning profiles of different pulse shapes was dealt with in the timing case by making a model of von Mises functions (see Section \ref{Sec:timing} and Fig \ref{Fig:model}). Seeing that the PA is often similar for all 3.1\,GHz observations, we have also aligned the profiles by using the value of ${\phi}_0$ that was estimated from the RVM model. The result is shown in Fig \ref{Fig:align}.
Comparing the two alignment methods, it again becomes obvious that the polarization from the magnetar is changing with time. At first glance it might seem like the aligning after the total intensity profile (the bottom plot of Fig. \ref{Fig:align}) is the one that should be preferred. However, a closer look at the PAs shows not only a shift in absolute value of the PAs but also that the PA curve for some of the observations have different slopes. For example, this can be seen by comparing the brown crosses with the dark blue ones in the upper panel of the bottom plot in Fig \ref{Fig:align}. Both PAs have similar slope in the leading edge of the pulse, but towards the trailing end the brown PA curve flattens out much faster than the dark blue one. 
The same slope difference is of course also present when we align the profiles after the value of $\phi_{0}$, but
by looking at the top plot of Fig \ref{Fig:align} (and disregarding from the light blue curve, which is offset from the others)  
even with the different slopes we find that the PAs align fairly well for the different observations. The profiles however do not. 

If the rotation measure (RM) of the interstellar medium would change with time between observations, that could explain the offset in absolute value of the PAs in the bottom panel of Fig. \ref{Fig:align}. However, by estimating the RM value for two of the observations where the offset is large (e.g. for the dark blue and the green curves) we do not see a large enough change between our observations. 

It is hard to say which (if any) of the two alignment methods describes the true magnetar emission. One way to get a better handle on this would be to use the polarization information in the timing of the pulsar by calculating TOAs using the $\phi_0$ values from each observation. The errors from this timing analysis could then be compared to the errors from the timing model described in Sec. \ref{Sec:timing}. Such an analysis would only be possible to do with 3.1\,GHz (or higher frequency) data, since part of the linear polarization emission at 1.4\,GHz is affected by interstellar scattering (see Section \ref{Sec:depolarization}), 
which causes the degree of linear polarization for many of the 1.4\,GHz observations to be too small or the PA swing to be flattened, preventing a reliable RVM fit. 
The integrated magnetar emission also seems to be more stable at 3.1\,GHz than it does at 1.4\,GHz, which would help in the timing for both timing methods. Unfortunately the number of 3.1\,GHz observations carried out at this point is not large enough and the observations that exist are not sampled densely enough to allow for such an analysis at the present time.

\subsection{Depolarization}
\label{Sec:depolarization}
In general, we observe a lower degree of linear polarization at 1.4\,GHz than at the higher observing frequencies. A similar trend is seen in 1E\,1547.0--5408 by \cite{CamApJ679}. 
This magnetar is at a similar DM as PSR\,J1622--4950 and they are both positioned close to the Galactic disk, which implies that both pulsars have a fairly high scattering timescale as predicted by the NE2001 model \citep{cor02}: 1E\,1547.0-5408 has DM = 830 $\pm$ 50 cm$^{-3}$pc and (l,b) = (327.23, --0.13) \citep{CamApJ666} which gives a scattering time scale of $\sim$70 ms at 1\,GHz compared to DM = 820 $\pm$ 30 cm$^{-3}$pc at (l,b) = (333.85, --0.10) for PSR\,J1622--4950 \citep{lev10} which results in $\sim$95 ms at the same frequency. 
\cite{CamApJ679} explain this depolarization in 1E\,1547.0--5408 at lower observing frequencies partly as an effect of the interstellar scattering of the pulse profile at these frequencies. This causes the PA to rotate through the different phases of the pulse, which will be mixed at the observer and will thus reduce the apparent linear polarization. They also give deviations in rotation measure (RM) over different paths taken by the scattered rays as a further possible reduction effect. They conclude that scattering effects can only be responsible for part of the depolarization. Hence, even though some of the depolarization at lower frequencies for PSR\,J1622--4950 could be intrinsic to the source, it is likely that scattering effects are also responsible for some depolarization in our case. 
In addition, we do observe a large amount of scattering in the single pulses at 1.4\,GHz (see Sec. \ref{Sec:scattering}), which further justifies this hypothesis. 
Unfortunately, due to issues with the observing system at the time of collection of the 1.4\,GHz single pulse data, we are not able to polarization calibrate this data, and hence will not be able to analyse which effect the depolarization has on the 1.4\,GHz single pulses. The single pulse data collected at an observing frequency of 3.1\,GHz is almost 100\% linearly polarized, which is discussed in Sec. \ref{Sec:SPchar} below.


\section{Single Pulses}
Single pulse analyses have been carried out at three observing frequencies, with bands centered at 1.4, 3.1 and 17\,GHz collected using the APSR backend at Parkes. 
Due to the large scattering effects at 1.4\,GHz (discussed in Sec. \ref{Sec:scattering} below) and the poor time resolution obtained at 17\,GHz (1024 bins over the pulse profile), most of the single pulse analysis carried out for this paper has focused on 3.1\,GHz data. 
In addition, one archival observation, collected within the frame work of the Methanol Multibeam Survey \citep{bat11} at Parkes using the Analogue Filterbank at a frequency band centered at 6.6\,GHz has been analysed.

\subsection{Scattering}
\label{Sec:scattering}
Due to the high DM of PSR\,J1622--4950, the single pulses are likely broadened enough by interstellar scattering at 1.4\,GHz that it will affect the pulse width at our time resolution. The pulse broadening due to scattering at the pulsar position and distance is $\sim$17\,ms at 1.4\,GHz according to the NE2001 model \citep{cor02}, assuming Kolmogorov scalings from 1.0\,GHz, but the large uncertainties in the model indicate that this value could deviate by up to a factor of ten \citep[as has been shown by e.g.][]{bha04}.
To better estimate the total scattering we analysed the widths of bright single pulses in the 3.1\,GHz data by dividing the observed frequency band up in parts and calculating the pulse broadening over the bandwidth for that observation. This resulted in $\sim$8.7\,ms smearing over a 768\,MHz band centered at 3.1\,GHz. Again by assuming Kolmogorov scaling ($\tau_{scatter} \propto \nu^{-\alpha}$, using scaling index $\alpha$ $\approx$ 4.0 as an estimation for high DM pulsars \citep{loh01,bha04}) we calculate a scattering of $\sim$200\,ms at a band centered at 1.4\,GHz, which is of the same order of magnitude as the measured value of the single pulse widths at that observing frequency (see Sec. \ref{Sec:SPwidth} and Tab. \ref{tab:width}). This implies that single pulses at 1.4\,GHz will be highly dominated by scattering effects and hence we will not include the 1.4\,GHz data in the single pulse analysis. The same analysis results in 0.41\,ms of smearing at 6.6\,GHz and 9.3\,$\mu$s at 17\,GHz.

\subsection{Single pulse characteristics}
\label{Sec:SPchar}

\begin{figure*}
	\begin{center}
		\includegraphics[width=17cm]{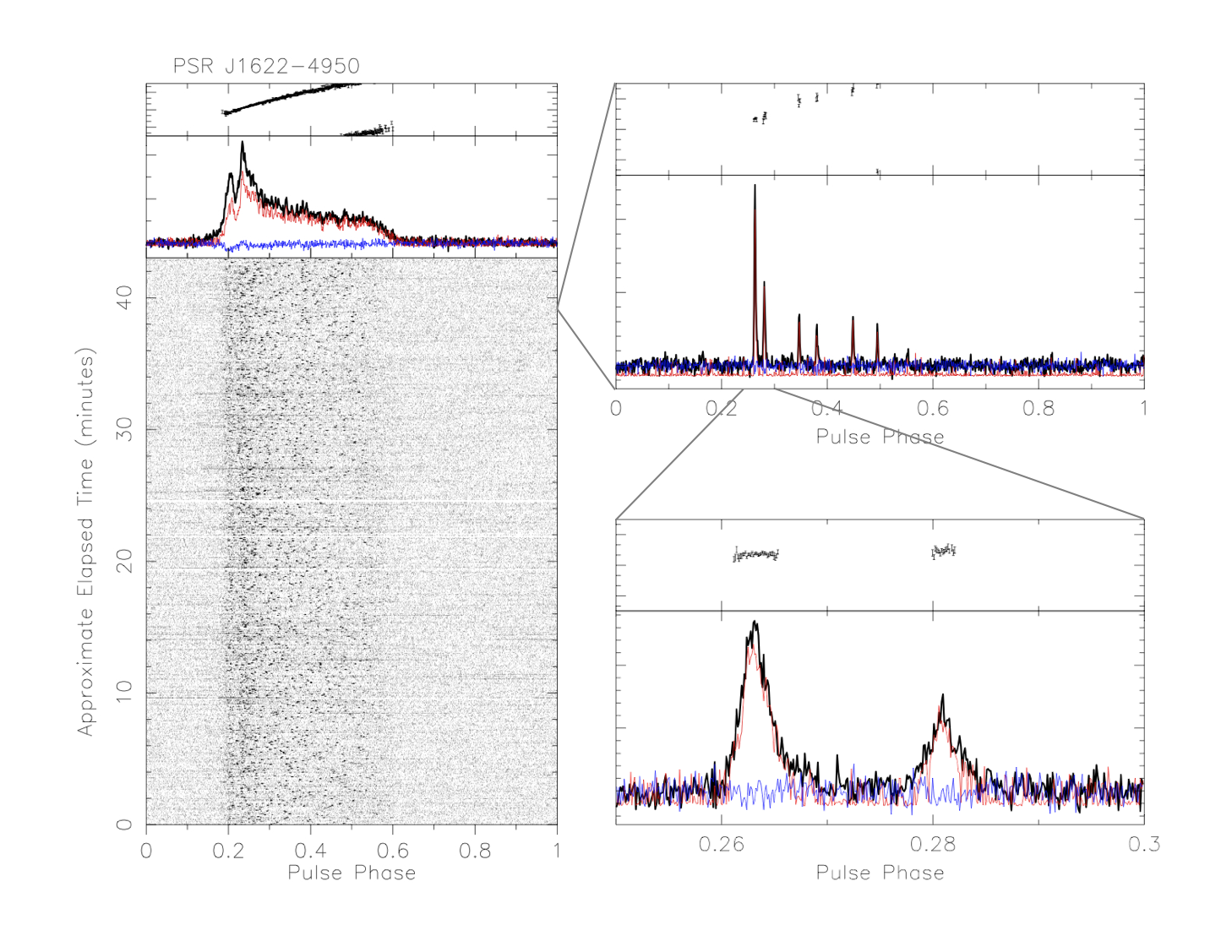}
	\caption{
	{\it Left:} 3.1\,GHz single pulse stack with the corresponding integrated profile on top. Linear polarization is shown in red, circular polarization in blue and the total intensity in black. 
	{\it Top right:} Pulse profile of one of the brighter single pulses showing the very narrow spikes of emission.
	{\it Bottom right:} Zoom in on the two brightest spikes of emission in the rotation above. 
	}
	\label{Fig:stack}
	\end{center}
\end{figure*}

Although the integrated pulse profile for PSR\,J1622--4950 is often very wide (with a $\sim$50$\%$ duty cycle), when analysing each rotation of the pulsar separately, it is clear that the profiles are built up by the collection of much narrower pulses (see Fig. \ref{Fig:stack} and Tab \ref{tab:width}). Each pulse consists of one or a few narrow spikes, at all observed frequency bands. 

The top right panel of Fig. \ref{Fig:stack} shows one of the brighter single pulses observed at 3.1\,GHz and the bottom right panel is a zoom-in on the two brightest components of that particular rotation. From this plot we can see how each spike is almost 100\% linearly polarized with no or very little circular polarization. The position angle of linear polarization of the separate components are well resolved and seem to be fairly flat across the spike. Looking at the combined PAs for the rotation in the top right panel, it is clear that they follow the PA swing of the total integrated profile. When we look at the single spike PAs in more detail, we find they are often in agreement with the total PA swing, but there are also occasions when the single PAs are much steeper than the integrated PA. This results in that we sometimes see small `wiggles' in the PA swing that often correspond to emission peaks in the total intensity of the integrated profile. Similar wiggles in the PA swing are also seen in some observations of  XTE\,J1810--197 \citep{kra07}.

\begin{figure}
	\begin{center}
		\includegraphics[width=8cm]{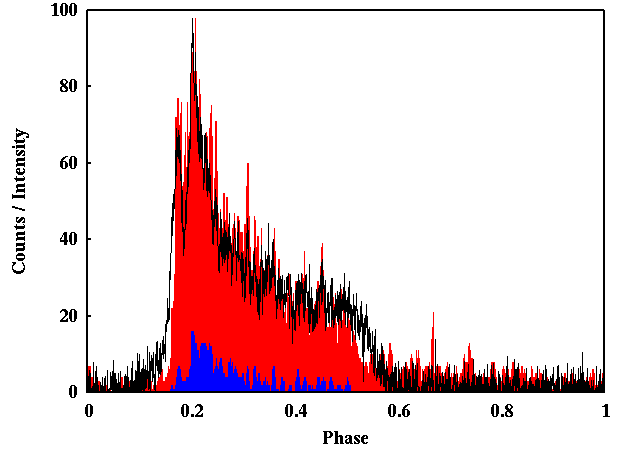}
	\caption{3.1\,GHz pulse phase histogram. The red boxes show all pulses, while the blue boxes only account for the brightest 10\% of the spikes. The black line is the total intensity profile averaged to match the top value of the histogram.}
	\label{Fig:phasehist}
	\end{center}
\end{figure}

At 3.1\,GHz a histogram of the phases of the single pulses matches the total intensity profile well (see Fig \ref{Fig:phasehist}). When only the brightest 10\% of the spikes are taken into account, we get the blue distribution in Fig. \ref{Fig:phasehist}. These spikes seem to be spread out over almost the entire pulse profile, but with a preference for certain phase bins. 
Each rotation consists of up to 15 spikes of emission, with 2 or 3 spikes per rotation being most frequent. 
The separation between emission spikes varies up to the full integrated pulse width, but with a preferred value of $\sim$170\,ms.

\subsection{Width of single emission spikes}
\label{Sec:SPwidth}
The width of the total integrated pulse profiles scales with observing frequency in ordinary pulsars. This phenomenon is usually thought to be due to radius-to-frequency mapping, which suggests that the emission at different frequencies originates from different altitudes above the polar cap \citep{cor78}. A study of the integrated profile width is not as useful for PSR\,J1622--4950 unless simultaneous multifrequency observations are considered, as the widths of the profiles are changing with time when the profiles vary.
However, in a similar way to the integrated profiles, we expect to see a decrease in the width of single emission spikes if the emission at different frequencies are emitted at different altitudes. 
Indeed, there have been indications that the pulse width is scaling down with increasing observing frequency in previous pulsar work (see e.g. \cite{kra02} and references therein). This is even more clearly demonstrated in PSR\,J1622--4950, see Tab. \ref{tab:width}. By comparing the average widths of the single pulse spikes at the three highest observed frequencies (3.1, 6.6 and 17\,GHz) and correcting for the broadening due to scattering, we calculate a frequency dependence for the width of the single emission spikes for this source as $\tau \propto \nu^{-0.62\pm0.12}$. \cite{kra02} carried out a similar analysis for the width of the micro-structure in PSR\,B1133+16 and found a much flatter dependence: $\tau_\mu \propto \nu^{-0.06\pm0.10}$.

\cite{kra02} also confirmed a relationship between micro-structure width and pulse period first suggested by \cite{cor79} as a linear dependence. The single emission spike widths for PSR\,J1622--4950 at similar observing frequencies do not fit this relation, but
the width at 17\,GHz ($\leq$8.44\,ms, an upper limit estimated to two time bins) is closer to the predicted value of 3.0\,ms. 
This may suggest that the scattering at the lower observing frequencies is underestimated, and that the 17\,GHz width is the true width of the single emission spikes. 
However, since the scattering at 3.1\,GHz is the measured value, we find it more likely that the width scaling with observing frequency is intrinsic, which indicates that perhaps separate relations are required for different frequency bands. 
Other reasons to why the emission from PSR\,J1622--4950 does not fit the relation could be that 
the pulse structure that we observe in PSR\,J1622--4950 is not of the same origin as microstructure, or that magnetars follow a different relation than ordinary pulsars. 
The single pulse widths for the other two radio-emitting magnetars are in the same order of magnitude as for PSR\,J1622--4950, and with the discovery of more sources it will be possible to investigate if a similar dependence for magnetars is valid. 

\begin{table}
\begin{center}
\caption{Widths of single pulse spikes at different observing frequencies. The upper limit given for the 17.0\,GHz pulse width is the value of two phase bins. The scatter broadening is measured at 3.1\,GHz and scaled using Kolmogorov scaling, with index $\alpha$ = 4.0, to estimate the values at the other frequency bands.}
\begin{tabular}{c | c c c}
\hline
\hline
{\bf Observing} & Average width & Scatter broadening & Width of \\ 
{\bf frequency} & of single pulses & of single pulses & integrated profiles\\
{\bf [GHz]} & [ms] & [ms] & [ms]\\
 \hline
{\bf 1.4} 	& 215                 	& 200 	& 2163 \\
{\bf 3.1}  	& 26 			& 8.4	  	& 1946 \\
{\bf 6.6}  	& 13	 		& 0.41   	& 1298 \\
{\bf 17.0} & $\leq$ 8.44 	& 0.0093  & 1730 \\
 \hline
 \hline
\end{tabular}
\label{tab:width}
\end{center}
\end{table}

\subsection{Pulse-flux distribution}
\begin{figure}
		\includegraphics[width=8cm]{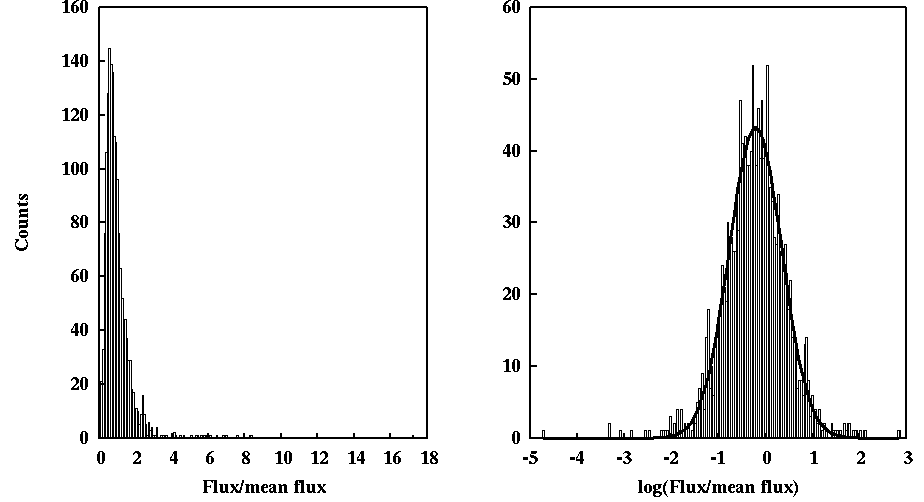}
		\caption{Pulse-flux distribution at 3.1\,GHz. 
		{\it Left:} Histogram of flux/mean flux of the single pulse spikes.
		{\it Right:} Histogram of the logarithm for the same data. The solid line shows the best fit of a Gaussian to the histogram, displaying the lognormal distribution of the flux density. 
		}
	\label{Fig:10cmfluxhist}
\end{figure}

By looking at the flux density of single spikes of emission we have calculated a pulse flux distribution for the single pulse spikes at 3.1\,GHz. 
A conventional energy distribution calculation often only records the peak flux density in the on-pulse region for each rotation and compares the resulting histogram to a histogram of the peak flux density in the off-pulse region. To account for the frequent occurrence of multiple emission spikes in a single rotation of PSR\,J1622--4950 we have approached this analysis differently. 
The on-pulse phases were chosen as all spikes that consisted of at least 3 consecutive phase bins with a signal stronger than 3 sigma, and the corresponding flux value for each spike was added to the histogram. 
The histogram in Fig. \ref{Fig:10cmfluxhist} was created after adding the phase bins to a total of 512 bins over the profile. The left panel shows the flux distribution and the right panel shows the 10 base logarithm of the flux. From this it is clear that the flux distribution from the magnetar is closer to a log-normal rather than Gaussian distributed. It is also evident from these plots that most spikes are of similar flux density and that there were no giant pulses observed from the magnetar. 
These properties are similar to the general ordinary pulsar population, for which a large fraction of the measured flux density distributions are close to log-normal and most do not emit giant pulses \citep[e.g.][]{cai04,bur11}.

\subsection{Pulse modulation}

To get an overview of to which extent the pulses are varying over the pulse phase, we have calculated two values for each bin in the pulse profile: the modulation index and the R parameter. We define the modulation index as $m_i$ = $\sigma_i/\mu_i$, where 
$\sigma_i$ is the mean intensity in bin $i$ and $\mu_i$ is the standard deviation in the $i$th bin after integrating over the entire observation. While the modulation index is a good indicator of whether there are persistent oscillations within the pulsed emission (such as e.g. drifting sub-pulses), it is not very sensitive to infrequent bursts of emission. In addition, it is hard to measure in observations with a low signal to noise value. To account also for non-persistent signal variation, we have calculated the R-parameter, $R_i= ({\rm MAX}_i - \mu_i)/\sigma_i$, (where MAX$_i$ is the maximum intensity in the $i$th bin) as is described in \cite{joh01}.  $R_i$ indicates the presence of very bright, infrequent spikes of emission in the separate pulse bins. The off-pulse value for the R-parameter will increase with the number of rotations of the pulsar due to Gaussian noise statistics, while the modulation index is undefined in the off-pulse region.

\begin{figure}
	\begin{center}
		\includegraphics[width=8cm]{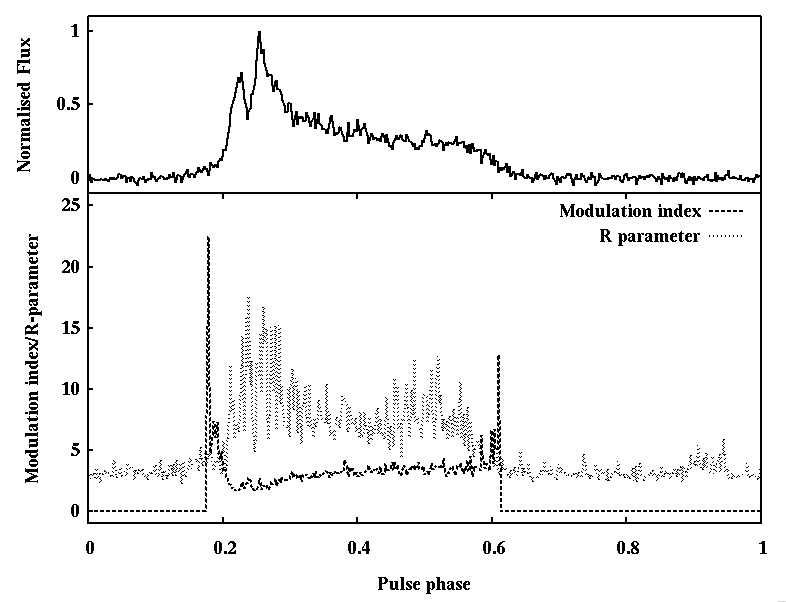}
	\caption{The bottom plot shows the modulation index (dashed line) and the R-parameter (dotted line). For reference, the integrated pulse profile is plotted in the top panel.}
	\label{Fig:10cmR}
	\end{center}
\end{figure}

Fig. \ref{Fig:10cmR} shows how the modulation index and the R-parameter is varying over the pulse profile for the 3.1\,GHz single pulse observation. 
The minimum and maximum values of the modulation index for bins which are clearly within the on-pulse region are $m_{min}$ = 1.7 and $m_{max}$ = 4.5. These values are in the upper range of values for ordinary pulsars \citep{wel05, bur11}.
From the R-parameter we can clearly see that the burst modulation is greater in the leading and the trailing edge of the on-pulse region than it is in the middle of the pulse. 
Overall the R-parameter is high for this source, as well being among the upper range of measured values for a larger pulsar sample \citep{bur11}. This further demonstrates the irregular spikiness of the emission at each rotation. 
Also in the case of XTE\,J1810--197 the integrated pulse profile consists of strong spiky sub-pulses \citep{ser09}. However, some of the sub-pulses for XTE\,J1810--197 could be considered giant pulses, but with broader pulse widths. \cite{ser09} also report on a modulation index for the single pulses that is high on average and that increases with increasing observing frequency but varies between components also within the same observing frequency. 
Similar spiky emission has been observed also in ordinary pulsars, such as B0656+14 \citep{wel06}. Again in contrast to PSR\,J1622--4950, this source also emits giant pulses and here the stronger pulses appear spikier than the weaker ones. 

\section{Conclusions}

The peak flux density of PSR\,J1622--4950 has decreased by a factor of  $\sim$2 since the discovery, and analyses of non-pulsed observations confirm this trend \citep{and11b}.
The timing analysis of the source implies large variations in the rate of spin-down. We find a correlation between flux density decrease and spin-down decrease, but are cautious not to over-interpret this result.  

The polarization is changing greatly between observations, but in general we observe a high degree of linear polarization and low circular polarization at frequencies higher than 1.4\,GHz. The observations at 1.4\,GHz are highly affected by interstellar scattering, which causes depolarization in the linear component, but not all changes in the linear polarization at this frequency can be due to propagation effects. We also see orthogonal phase jumps, flat PAs and changed handedness in the circular polarization in some of the observations at 1.4\,GHz and also at higher observing frequencies. 
RVM fits imply that the geometry of PSR\,J1622--4950 is nearly aligned, with the line of sight remaining within the emission beam for large parts of the rotation. 
If this proves to be a general property for magnetars, it could provide insight in to why so few of the magnetars have observed radio pulsations. 

The single pulses from the magnetar are very narrow in comparison to the width of the total integrated profile. The emission from each rotation consists of a few narrow spikes, that vary in longitude and separation between pulses. The widths of the single emission spikes appear to scale inversely with observed frequency, by getting narrower as the frequency gets higher. This could demonstrate that the emission spikes observed at different frequencies are emitted at different altitudes above the polar cap in the neutron star. 
The pulse flux distribution points towards a log-normal flux distribution of the spikes, without any signs of giant pulses from the magnetar. The spikiness in the emission results in a very high R-parameter across the pulse profile and a modulation index that is slowly increasing with pulse longitude and is higher in the pulse edges. 

In all, PSR\,J1622--4950 has many properties that are very similar to the other radio magnetars, but also some things that differ, such as variations in linear polarization and PA swing on short time scales and a non-smooth frequency derivative evolution. 
We stress the importance of regular monitoring of these sources as the known sample is very small and the only chance to understand their emission and their connection to other neutron stars is by continuing to observe their various properties. 
In particular, it is important to establish whether the overall flux density is the only property that is changing on longer time scales, or if also other emission properties are varying. 
In order to discover more radio magnetars we need to understand the time scales of their on and off periods, and depending on their special features, we may need to re-evalute the way searches for these sources should be performed.  

\section{Acknowledgements}
We wish to thank M. Livingstone for suggesting the glitch hypothesis in the timing section of this paper.
The Parkes Observatory is part of the Australia Telescope, which is funded by the Commonwealth of Australia for operation as a National Facility managed by CSIRO.

\end{document}